\documentclass[twocolumn,showpacs,preprintnumbers,amsmath,amssymb]{revtex4}
\usepackage{graphicx}% Include figure files
\usepackage{dcolumn}% Align table columns on decimal point
\usepackage{bm}% bold math
\begin{document} 

\title{Comment on `Non-existence of the Luttinger-Ward functional and
misleading convergence of skeleton diagrammatic series for Hubbard-like models'}
\author{R. Eder}
\affiliation{Karlsruhe Institute of Technology,
Institut f\"ur Festk\"orperphysik, 76021 Karlsruhe, Germany}
\date{\today}

\begin{abstract}
The known analytical properties of the Green's function and
self-energy rule out an ambiguity of the self-energy. The
noninteracting Green's functions obtained by Kozik {\em et al.}
in arxive:1407.5687 likely have pathological properties.
\end{abstract} 

\maketitle
In a recent publication\cite{Kozik} Kozik {\em et al.} 
studied the problem of reconstructing the noninteracting Green's function
$G_0(\omega)$ and self-energy $\Sigma(\omega)$ of various models from the 
known exact Green's function $G(\omega)$. The method employed
was to start from a given `trial $G_0(\omega)$' input this
into a Quantum Monte Carlo solver, compute the resulting 
$G[G_0](\omega)$ and enforce
$G[G_0](\omega)=G(\omega)$ by iterative improvement of $G_0$.
They found that depending on details of the iterative procedure
they could obtain different pairs $(G_0,\Sigma)$ which give
the same $G$, whereby $\Sigma$ is the (numerically) exact
self-energy obtained by the solver for the noninteracting
Green's function $G_0$. They concluded, that no unique mapping
$G\rightarrow \Sigma$ exists so that the Luttinger-Ward functional 
$\Phi[G]$ is non-existent.\\
However, under the physically reasonable assumption that $G_0^{-1}(\omega)$ 
be analytical the known analytical forms of the Green's function
and self-energy rule out any ambiguity of $\Sigma$ and hence $G_0$.\\
Luttinger\cite{ Luttinger_analytical} derived the following 
form of the self-energy: 
\begin{eqnarray}
\Sigma(\omega) = \eta + \sum_{i=1}^n\;\frac{\sigma_i}{\omega-\zeta_i}
\label{lutty}
\end{eqnarray}
with real $\eta$, $\sigma_i>0$ and $\zeta_i$.
Equation (\ref{lutty}) follows directly from the Lehmann-representation
for Green's function
\begin{eqnarray}
G(\omega) = \sum_{i=1}^{n+1}\;\frac{Z_i}{\omega-\omega_i}
\label{greeny}
\end{eqnarray}
where $Z_i>0$ and $\omega_i$ are real. 
On the real axis, $G(\omega)$ crosses zero once
in each interval $[\omega_i:\omega_{i+1}]$ with negative slope.
Assuming $G_0^{-1}$ to be analytical this implies
a divergence of $\Sigma(\omega)$, hence the $\zeta_i$ equal the
zeroes of $G(\omega)$
and the number of poles of $\Sigma(\omega)$
is one less than the number of poles of $G(\omega)$
(all of this is well-known in the literature).
For analytical $G_o^{-1}(\omega)$ one has near a given $\zeta_i$
\[
G^{-1}(\omega) = -\frac{\sigma_i}{\omega-\zeta_i} + c_0 + c_1(\omega-\zeta_i)
+\dots
\]
with real constants $c_0$ and $c_1$ and omitted terms
are of higer order in $\omega-\zeta_i$,
so that
\begin{eqnarray*}
G(\omega) &=& -\frac{\omega-\zeta_i}{\sigma_i} - \frac{c_0}{\sigma_i^2}
(\omega-\zeta_i)^2 +\dots,\nonumber \\
\frac{d G(\omega)}{d\omega}|_{\omega=\zeta_i}&=&-\frac{1}{\sigma_i}.
\end{eqnarray*}
Under the assumption that $G_0^{-1}$ be analytical
all $\sigma_i$ and $\zeta_i$ thus are completely determined by
the zeroes of $G(\omega)$ and its slope at these.
The additive constant $\eta$ is the Hartee-Fock potential\cite{selfie}
which can be expressed as a functional of $G(\omega)$ and thus is
determined uniquely by $G(\omega)$ as well. 
One might worry about an `information mismatch' because
$G$ involves $2n+2$ parameters whereas $\Sigma$ involves only
$2n+1$ - however, one constant in $G$ is absorbed by
the requirement $\sum_i Z_i=1$.
To summarize the discussion so far: if one imposes the rigorously
known analytic structure of $G(\omega)$ and $\Sigma(\omega)$
and the analyticity of $G_0^{-1}(\omega)=\omega + \mu - H_0$
there is exactly one self-energy
and - by the Dyson equation - exactly one $G_0^{-1}$
for each physical $G(\omega)$ (whereby for a physical $G(\omega)$ 
the equation $G_0^{-1}(\omega) -\Sigma(\omega)=0$ must be obeyed
for all $\omega_i$).\\
The above discussion suggests that the solutions
obtained by Kozik {\em et al.}
have a nonanalytical $G_0^{-1}(\omega)$
and this indeed becomes apparent in their Figure 3, where
for $U=4$ the self-energy for the unphysical solution
does not diverge as $\omega\rightarrow 0$.
One the other hand one knows that
\begin{eqnarray*}
G^{-1}(\omega) &=& G_0^{-1}(\omega) - \Sigma(\omega)\nonumber \\
&=&-\frac{U^2}{4\omega}\left(1-\frac{4\omega^2}{U^2}\right)
\end{eqnarray*}
so if $\Sigma(\omega)$ stays regular at 
$\omega\rightarrow 0$ necessarily 
$G_0^{-1}(\omega) \rightarrow -\frac{U^2}{4\omega}$. To see the consequence
let us assume that the regular part of $G_0^{-1}(\omega)$ has the simplest form
compatible with particle-hole symmetry and the requirement
that $G_0(\omega)\rightarrow 1/\omega$ for large $|\omega|$ whence
\begin{eqnarray*}
G_0(\omega)=\frac{1}{\omega -\frac{U^2}{4\omega}},
\end{eqnarray*}
which would be the exact, fully interacting Green's function.
This suggests that the solutions obtained by  Kozik {\em et al.} 
correspond to a `noninteracting Green's function' which up
to slight deviations is equal to the fully interacting one
and a small `self energy' which undoes these slight deviations.\\
While in the absence of precise information on the
$G_0$ found by Kozik {\em et al.} this is somewhat speculative
it should be noted that the fact that these authors may be using
a slightly modified fully interacting Green's function 
as their noninteracting Green's function, might provide a rather natural
explanation for the `misleading convergence of the skeleton series'.
Namely in their Figure 3c
Kozik {\em et al.} are comparing self-energies obtained
in two different ways:\\
1) A `noninteracting Green's function' which likely is close to the
exact interacting one and in particular already has the two Hubbard bands,
is used as input in a quantum Monte Carlo solver and
a self-energy is obtained, which necessarily can only be a small
correction anymore.\\
2) The exact interacting Green's function which also has the two
Hubbard bands is inserted into a low order skeleton-diagram
expansion for the self-energy. Since particle-hole excitations
have to cross the Hubbard gap, the resulting self-energy also
must be small.\\
That the self-energies obtained in this way show some similarity then
may not be surprising.

\end{document}